\documentclass{article}
\usepackage{authblk}
\usepackage{amsmath}
\usepackage{amsfonts,amssymb,mathrsfs,graphicx}
\usepackage[style=alphabetic]{biblatex}
\usepackage{multirow}
\author[1]{Jayati Kaushik\thanks{jayati@edufor.me}}
\author[2]{Aaruni Kaushik}
\author[3]{Upasana Parashar}

\affil[1,3]{Independent Researcher}
\affil[3]{RPTU}

    \makeatletter
    \renewcommand\AB@affilsepx{: \protect\Affilfont}
    \makeatother

    \affil[ ]{Email ids}

    \makeatletter
    \renewcommand\AB@affilsepx{, \protect\Affilfont}
    \makeatother

    \affil[1]{jayati@edufor.me}

\title{Using Topological Data Analysis to classify Encrypted Bits}
\date{}
\providecommand{\keywords}[1]
{
  \small	
  \textbf{\textit{Keywords---}} #1
}
\begin{document}
    \maketitle
    \begin{abstract}
        We present a way to apply topological data analysis for classifying encrypted bits into distinct classes. 
Persistent homology is applied to generate topological features of a point cloud
obtained from sets of encryptions. We see that this machine learning pipeline is able to classify our data successfully
where classical models of machine learning fail to perform the task. We also see that this pipeline works 
as a dimensionality reduction method making this approach to classify encrypted data a realistic method to classify the given encryptioned bits
    \end{abstract}
    \keywords{Topological data analysis, machine learning, cryptography, GSW 13, supervised machine learning, classification, persistent homology, filtration}
    \section{Introduction}\label{sec1}

Topological Data Analysis (TDA) uses techniques from algebraic topology to study and extract the geometric and topological features on the shape of data. TDA is very sensitive to patterns of both large and small scales that often stay undetected by other methods such as Principle Component Analysis and Cluster Analysis. TDA is also more robust to noise in the data\cite{TDAinsights}. 

In this paper we aim to use the powers of TDA to classify data that has been encrypted according to GSW 13 scheme \cite{cryptoeprint:2013/340}. Using this scheme, 0 and 1 was encrypted for different values of the parameter $n$ which gives approximately $n^2\times n^2$ matrix with binary entries. This is an asymmetrical cryptosystem therefore the Chosen Plaintext Attack (CPA) is trivial to mount, as we can access an oracle which encrypts the inputs for us. This opens up the possibility of applying supervised machine learning models on the encrypted bits in order to classify the data. The problem is then a two class classification problem for supervised machine learning. We encrypted 0 and 1, and we attempt to classify the encrypted bits.

The encrypted data consists of square matrices with binary entries. For classification many classical supervised machine learning algorithms; like Neural Networks, Tensor Classification, AutoKeras, Random Forest (on un-transformed data); were tested. All of these models failed to perfom the task. 

We treat the encrypted data as a binary images. We take inspiration from \cite{https://doi.org/10.48550/arxiv.1910.08345} and create a point cloud and apply a modified version of their pipline to our data. We see that TDA is able to not only successfully classify the data with great accuracy  but also acts as a potent dimensionality reduction technique; in one case reducing the number of features from $961^2$ to 55. This considerably reduces the compute time for our pipeline and makes this attack on the cryptosyttem realistic.

\section{Methodology}
\subsection{Persistent Homology on Point Clouds}
 Persistent homology extracts the births and deaths of topological features throughout a filtration built from a dataset. We  present the necessary definitions here. 
 The definitions in this sections are taken from  \cite{https://doi.org/10.48550/arxiv.1910.08345} and \cite{topinbio}
 \subsubsection{Simplicial Complexes}
 The \textit{k-simplex} spanned by the points $\{x_0,x_1,\dots, x_n\}$ is the set of all point 
 \begin{equation}\label{k-simplex}
     z = \sum_{i=0}^k a_ix_i, \hspace{0.5cm} \sum_{i=0}^k a_i=1
 \end{equation}
 For a given $z$ we refer to $a_i$ as the \textit{$i$th barycentric coordinate}  For example, a $0$-simplex is a point, 
 a $1$-simplex is a line segment with endpoints $x_0$ and $x_1$.\\
 For a simplex $S$ spanned by the points $P = \{x_0,\dots , x_n\}$, a \textit{face} of $S$ refers to any simplex spanned by a 
 subset of $P$. \\
 A \textit{simplicial complex} $X$ in $\mathbb{R}^n$ is a set of simplices in $\mathbb{R}^n$ such that 
 \begin{enumerate}
     \item Every face of a simplex in $X$ is also a simplex in $X$
     \item The interesction of two simplices in $X$ is a face of each of them.
 \end{enumerate}
A \textit{finite simplicial complex} is a simplicial complex with finitely many simplices.\\

 A simplicial complex can be obtained from a dataset using the Vietoris-Rips construction.\\
 Let $(X,\partial_X)$ be a finite metric space and fix $\epsilon>0$. The \textit{Vietoris-Rips complex} $VR_\epsilon(X,\partial_X)$ is the abstract simplicial complex with 
 \begin{enumerate}
     \item Vertices the points of $X$
     \item a $k$-simplex $[v_0,v_1,\dots v_n]$ when 
$$ \partial_X(v_i,v_j)\leq2\epsilon \hspace{0.5cm} \forall \hspace{0.5cm} 0\leq i,j \leq k$$
 \end{enumerate}
\subsubsection{Filtrations of Simplicial Complexes}
A \textit{filtration} of simplicial complexes is a nested sequence of simplicial complexes $$\{VR_\epsilon(X,\partial_X)\}_{\epsilon>0}$$
satisfying $VR_{\epsilon_1}(X,\partial_X) \subseteq VR_{\epsilon_2}(X,\partial_X)$

\subsection{Persistent Homology in Images}
\subsubsection{Images as cubical complexes}
A $d$-\textit{dimensional image} is a map $$\mathcal{I}: \mathscr{I} \subseteq \mathbb{Z}^d \rightarrow \mathbb{R}^n$$ \textit{Voxel} is an element $v\in \mathscr{I}$\\
\textit{Intensity} and the value $\mathcal{I}(v)$.
With slight abuse of terminology, any subset $\mathscr{I}\subseteq \mathbb{Z}^d$ is referred to as an image. \\
A voxel is represented by a $d$-cube and all of its faces are added. Let the resulting cubical complex be $K$. Other ways of representing images as cubical complexes can be found in \cite{missingref}.\\
Define the following function: 
$$ \mathcal{I}^\prime = min_{\sigma face \hspace{0.1cm} of \hspace{0.1cm} \tau} \mathcal{I}(\tau)$$ 
Let $K$ be the cubical complex built from image $\mathscr{I}$. Then, the $i$-th \textit{sublevel} of $K$ is given by
$$K_i = \{ \sigma \in K \\vert \mathcal{I}^\prime (\sigma)\leq i\} $$
The filtration of cubical complexes is then defined by the set $\{K_i\}_{i\Im(\mathscr{I}}$, indexed by the function $\mathcal{I}$
\subsubsection{Filtrations of Binary Images}
A \textit{Binary image} is defined as $\mathcal{B} : \mathscr{I}\subseteq \mathbb{Z}^d \rightarrow \{0,1\}$.\\
By building filtrations from it, the topological features of the binary image can be highlighted and extracted. \\
Many techniques for these filtrations exist. See \cite{https://doi.org/10.48550/arxiv.1910.08345} for details. Here, we define the two filtrations used in the final model. 
\begin{enumerate}
    \item \textit{Height Filtration}\\
    To define the height filtration $\mathcal{H} : \mathscr{I} \rightarrow \mathbb{R}$ 
    of a $d$-dimensional binary image $\mathscr{I}$ choose a vector $v \in \mathbb{R}^d$ of norm 1. Then $\forall p \in \mathscr{I}$, if $\mathcal{B}(p)=1$ then define $\mathcal{H}(p):= <p,v>$, the distance of $p$ to the hyperplane defined by $v$. If $\mathcal{B}(p)= 0$
    then $\mathcal{H}(p):= \mathcal{H}_\infty$, where $\mathcal{H}_\infty$ is the filtration value of the pixel that is farther from the defined hyperplane. 
    \item \textit{Radial Filtration}\\
    To define the radial filtration of $\mathscr{I}$ with centre $c\in \mathscr{I}$, assign to a voxel $p$ the value $\mathcal{R}(p) := ||c-p||$ if $\mathcal{B}(p)=1$ and 
    $\mathcal{R}_\infty$ if $\mathcal{B}(p)=0$, where $\mathcal{B}_\infty$ is the distance of the pixel farthest away from the center. 
\end{enumerate}

\subsection{Metrics and Kernels}
Various topological metrics used in this work are defined here. For detailed explanation of these, and how they are used in a machine learning pipeline, see \cite{cryptoeprint:2013/340} and \cite{topinbio}.\\
\begin{enumerate}

\item \textit{Amplitude} of a persistence diagram is its distance to the empty diagram which consists only points on the diagonal. This can be obtained using either $L1$ or $L2$ norms. \\
\item The \textit{Betti Curve} $B_n: \mathscr{I}\rightarrow \mathbb{R}$ of a barcode (see \cite{topinbio} for explanation) $D = \{(b_j,d_j)\}_{j\in \mathcal{I}}$ is a function that returns for each step $i \in \mathcal{I}$, number of bars $(b_j,d_j)$ that contain $i$.\\
\item \textit{Heat Kernel} is obtained by placing Gaussians of standard deviation $\sigma$ over every point of the persistence diagram and it's negative in the mirror image of the points across the diagonal. The output is a real valued function. \\
\item \textit{Wasserstein amplitude} of order $p$ is the $L_p$ norm of the point of point distances to the diagnial given by:
$$A_W = \frac{\sqrt{2}}{2}(\sum_i(d_i0b_i)^p)^{\frac{1}{p}}$$
\item \textit{Bottleneck amplitude} is obtained by letting  $p$ go to $\infty$ in the definition of wasserstein apmlitude above. This gives: $$A_B = \frac{\sqrt{2}}{2} sup_i(d_i-b_i)$$
\item \textit{Persistence entropy} of a barcode is calculated as follows: 
$$ PE(D) = \sum_{i=1}^n\frac{l_i}{L(B)}log(\frac{l_i}{L(B)}) $$
where = $l_i:= d_i-b_i$ and $L(B):=l_1+\dots+l_n$
\end{enumerate}
\subsection{Machine Learning Models}
In this work, we have used decision tree classifier and random forest classifier. \cite{MLProb} can be used for an explanation for both the classifiers.

The machine learning pipeline used is modified from \cite{https://doi.org/10.48550/arxiv.1910.08345} to fit our use case. Final parameters are recorded in the results section.


\section{Results and Discussion}
In this section we present our final results of our classification pipline. It was implemented using Giotto-tda library in python (https://giotto-ai.github.io/gtda-docs/latest/index.html). 

Once the pipeline was built and executed for the TDA, we used Random Forest Classifier and Decision Tree Classifier for the classification. The train test split was 0.7 and 0.3 respectively. Using Grid Search, it was determined that the best results are gotten when we do the filtration along the direction vector $(-1,1)$ and the best center for the radial filtration was determined to be $(13,13)$, for $n=6$. The direction has not changed with $n$. The filtration used were Height and Radial Filtrations. Cubical persistence was calculated for the persistence diagrams. Persistence Entropy was calculated for the vectorization. Both Random Forest and Decision Tree gave comparable results. These are summarized in the table below (with higher accuracy chosen):
\begin{figure}[h]
    \centering
    \includegraphics{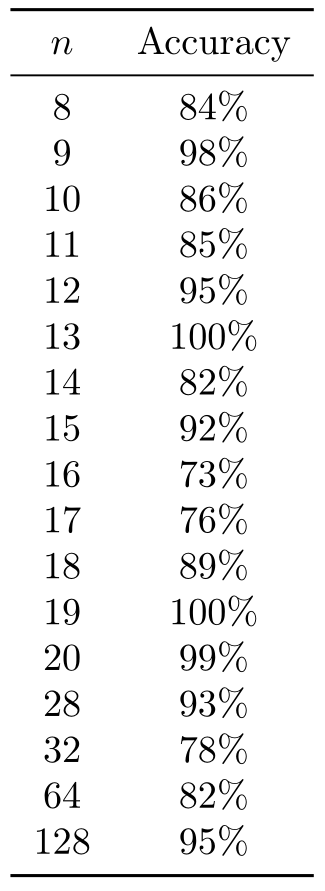}
    \caption{Accuracy achieved for various values of $n$}
    \label{fig:my_label}
\end{figure}

A comparison of both the classifiers is done below:
\begin{table}[h]
    \centering
    \begin{tabular}{c|c|c}
    
    n & Random Forest & Decision Tree \\
    \hline
    \hline
         28 &  .84 &  .93 \\
32 &  .78 &  .76\\
 64 &  .68 &  .82 \\
 128 &  .78 &  .95 \\
 
    \end{tabular}
    \caption{Comparison of accuracies of Decision Tree and Random Forest }
    \label{tab:my_label}
\end{table}

Since this is generated data, and classes are completely balanced and sampling is stratified, accuracy is sufficient to assess model performance.
We are yet to check the feature importance. That might give us further insight into the explainability of the model. We are also yet to see if this keeps working for very large values of $n$. It would also be of interest to understand why Decision Tree seems to perform better most of the time.
\section*{Acknowledgments} 

\addcontentsline{toc}{section}{Acknowledgments} 
I thank my friends and family; for without their support this paper would not be possible. 
\printbibliography

\end{document}